\documentclass[twocolumn,prb,showpacs,prbprintnumbers,fleqn,superscriptaddress]{revtex4}
\usepackage{epsfig}
\usepackage{graphicx}
\usepackage{amsfonts}
\newcommand{\ct}{\cite}
\newcommand{\bi}{\bibitem}
\newcommand{\de}{\delta}

\newcommand{\ga}{\gamma}
\newcommand{\si}{\sigma}
\newcommand{\bra}{\langle}
\newcommand{\ket}{\rangle}
\newcommand{\non}{\nonumber}
\newcommand{\dg}{\dagger}
\newcommand{\be}{\begin{equation}}
\newcommand{\ee}{\end{equation}}
\newcommand{\ba}{\begin{eqnarray}}
\newcommand{\ea}{\end{eqnarray}}

\begin{document}

\title{Quenching along a gapless line: A different exponent for defect density}
\author{Uma Divakaran}
\email{udiva@iitk.ac.in}
\author{Amit Dutta}
\email{dutta@iitk.ac.in}
\affiliation{Department of Physics, Indian Institute of Technology Kanpur 
208 016, India}
\author{Diptiman Sen}
\email{diptiman@cts.iisc.ernet.in}
\affiliation{Center for High Energy Physics, Indian Institute of Science,
Bangalore 560 012, India}

\date{\today}

\begin{abstract}
We use a new quenching scheme to study the dynamics of a one-dimensional 
anisotropic $XY$ spin-1/2 chain in the presence of a transverse field which 
alternates between the values $h+\de$ and $h-\de$ from site to site.
In this quenching scheme, the parameter denoting the anisotropy of interaction
($\ga$) is linearly quenched from $-\infty$ to $ +\infty$ as $\ga = 
t/\tau$, keeping the total strength of interaction $J$ fixed. The system 
traverses through a gapless phase when $\ga$ is quenched along the critical
surface $h^2 = \de^2 + J^2$ in the parameter space spanned by $h$, $\de$
and $\ga$. By mapping to an equivalent two-level Landau-Zener problem, we 
show that the defect density in the final state scales as $1/\tau^{1/3}$, a 
behavior that has not been observed in previous studies of quenching through 
a gapless phase. We also generalize the model incorporating additional 
alternations in the anisotropy or in the strength of the interaction, and 
derive an identical result under a similar quenching. Based on the above 
results, we propose a general scaling of the defect density with the quenching
rate $\tau$ for quenching along a gapless critical line.
\end{abstract}

\maketitle

\section{introduction}

The dynamics of a quantum system swept across a quantum critical point at a 
uniform rate has been studied extensively in recent years. Since a quantum 
phase transition \ct{sachdev99,dutta96} is necessarily accompanied by a 
diverging correlation length as well as a diverging relaxation time, the 
dynamics of the system cannot be adiabatic for the entire period of the 
evolution however slow the variation in the parameter may be. (The relaxation
time of a quantum system is given by the inverse of the energy gap which goes
to zero at the quantum critical point). Assuming that the system was
initially prepared in its ground state, the non-adiabaticity near a quantum 
critical point prevents the system from following its instantaneous ground 
state resulting in the production of defects in the final state. 

The Kibble-Zurek (KZ) argument \ct{kibble76} asserts that the non-adiabatic 
effect becomes prominent only close to the critical point when the rate of 
change of the Hamiltonian is of the order of the relaxation time of the 
underlying quantum system. When a parameter of the quantum Hamiltonian is 
varied as $t/\tau$, where $\tau$ is the characteristic time scale of the 
quenching, the above argument predicts a density of of defects in the final 
state that scales as $1/\tau^{d \nu /(z \nu +1)}$ in the limit of $\tau \to 
\infty$. Here $\nu$ and $z$ denote the correlation length and dynamical 
exponents, respectively, characterizing the associated quantum phase 
transition of the $d$-dimensional quantum system. The KZ prediction has been 
verified for various exactly solvable spin models when quenched across a 
critical \ct{zurek,levitov} or a multicritical \ct{mukherjee} point at a 
uniform linear rate. The above studies have been generalized to explore the 
defect production in a non-linear quench across a quantum critical point where
a parameter in the Hamiltonian is varied as $|t/\tau|^{\alpha}$ with $\alpha >
0$ \ct{sen}. Recent experimental studies on the dynamics of quantum systems 
\ct{expts}, especially quantum magnets \ct{wernsdorfer}, ultracold atoms 
trapped in optical lattices \ct{duan} and spin-one Bose-Einstein condensates 
\ct{sadler}, have paved the way for a plethora of related theoretical studies 
\ct{zurek,levitov,mukherjee,sen,sengupta04,dziarmaga06,mukherjee08,
patane08,dziarmaga08,polkovnikov08}.

Another interesting scenario emerges when a low-dimensional quantum system is 
quenched {\it through} a gapless phase or an extended quantum critical region
\ct{polkovnikov07,sengupta08,santoro08}. It has been established that when a 
$d$-dimensional system is quenched along a $(d-m)$-dimensional critical
surface, the scaling of the defect density with $\tau$ is modified to a 
generalized KZ form given by $1/\tau^{m \nu /(z \nu +1)}$ \ct{sengupta08}. 

In the present work, we explore the dynamics of a one-dimensional anisotropic 
$XY$ spin-1/2 chain in the presence of a transverse field which alternates 
between $h+\de$ and $h-\de$ from site to site. We employ a new quenching
scheme in which the parameter determining the anisotropy of interaction is 
quenched as $t/\tau$, keeping the strength of interaction fixed, in such a way
that the system is driven along a gapless line on a critical surface in the 
parameter space. We show that the density of defect scales as $1/\tau^{1/3}$, 
a result that cannot be explained by the previous studies on the quenching 
through a gapless phase. We also propose a general scaling relation for 
such a quenching dynamics along a gapless line.

The paper is organized as follows. Our model, the quenching scheme and the 
results obtained for the generation of defects are presented in Sec. II. At 
the end of that section, we propose a general scaling relation for the defect 
density when a system is quenched along a gapless line. We end with some 
concluding remarks in Sec. III.

\section{The quenching dynamics and the result}

The Hamiltonian of the spin-1/2 anisotropic $XY$ model with an alternating 
transverse field is given by \ct{perk,viola08}
\ba &H& =~ - ~\frac{1}{2} ~[ ~\sum_j ~{(J_x + J_y)} (\si^x_j \si^x_{j+1} +
\si^y_j \si^y_{j+1})~ + \non \\ 
&+& {(J_x-J_y)} (\si^x_j \si^x_{j+1} - \si^y_j \si^y_{j+1}) + 
(h-(-1)^j\de ) \si^z_j)], \label{h2} \ea
where $\si$'s denote the Pauli spin operators satisfying the standard 
commutation relations. The strength of the transverse field coupled to the 
operator $\si^z$ alternates between $h+\de$ and $h-\de$ on the odd and 
even sites respectively. We have chosen all the interactions and the fields 
to be non-random. Henceforth, we shall refer to $J_x+J_y =J$ as the strength 
and $\ga= J_x-J_y$ as the anisotropy of the nearest-neighbor interaction.

To map the spin operators to spinless fermion operators using the 
Jordan-Wigner transformation \ct{lieb,kogut}, we note that the presence of 
two underlying sub-lattices necessitates the introduction of a pair of fermion
operators $a$ and $b$ \ct{perk,viola08} for even and odd sites as defined below
\ba \si_{2j}^+ &=& b_{2j}^\dg ~\exp[i\pi\sum_{l=1}^{j-1}
b_{2l}^\dg b_{2l} + i\pi \sum_{l=1}^ja_{2l-1}^\dg a_{2l-1}], \non \\
\si_{2j+1}^+ &=& a_{2j+1}^\dg ~\exp[i\pi\sum_{l=1}^{j}
b_{2l}^\dg b_{2l}+i\pi\sum_{l=0}^{j-1}a_{2l+1}^\dg a_{2l+1}]. \label{jw} \ea
Using a restricted zone scheme (where the wave vector $k$ ranges from 
$-\pi/2$ to $\pi/2$) in the Fourier space, the Hamiltonian can be written as
$$H ~=~ \sum_k ~H_k ~=~ \sum_k ~{\hat A^\dg}_k ~\hat H_k ~{\hat A}_k,$$ where
$\hat A_k$ is the column $(a_k^\dg ,a_{-k},b_k^\dg, b_{-k})$. The $4\times 4$
Hermitian matrix $H_k$ is given by
\ba \left[ \begin{array}{cccc} h +J \cos k & i\ga \sin k & 0 & -\de \\
-i \ga \sin k & - h - J \cos k & \de & 0 \\
0 & \de & J \cos k - h & i \ga \sin k \\
-\de & 0 & - i \ga \sin k & - J \cos k + h \end{array} \right]. \non \\
& & \label{ham1} \ea
The excitation spectrum of the Hamiltonian $H$ is now obtained by 
diagonalizing the reduced Hamiltonian matrix $H_k$ and is given by
\ba \Lambda_k^\pm &=& [ h^2 +\de^2 +J^2 \cos^2 k + \ga^2 \sin^2 k \non \\
& & \pm 2 \sqrt{h^2 \de^2 + h^2 J^2 \cos^2 k + \de^2 \ga^2 \sin^2 
k}]^{1/2}. \label{lam3} \ea
Denoting the four eigenvalues by $\pm \Lambda_k^{\pm}$, we can write the 
spectrum of the Hamiltonian in the form
\be H ~=~ \sum_{-\pi /2 < k < \pi /2} ~~\sum_{\nu=+,-} ~\Lambda_k^\nu ~(
\eta_{k,\nu}^\dg \eta_{k,\nu} -\frac{1}{2}), \ee
where $\eta_{k,\nu}^\dg$ is the quasiparticle creation operator corresponding
to the mode $(k,\nu)$. In the ground state the levels $-\Lambda_k^+$ and $-
\Lambda_k^-$ are filled. At the quantum critical point, $\Lambda_k^-$ vanishes
at an ordering wave vector, and the critical exponents are obtained by 
studying the behavior of $\Lambda_k^-$ in the vicinity of the critical point.

\begin{figure}
\includegraphics[height=2.2in]{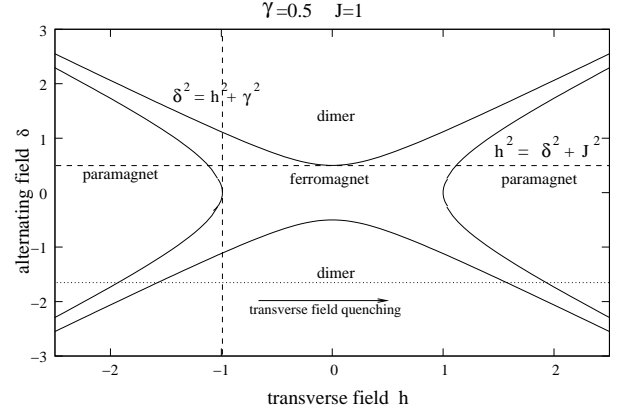}
\caption{Phase diagram of the $XY$ chain in an alternating transverse field. 
We have chosen $J=1$ and the critical lines are drawn in the $\ga=0.5$ 
plane. Two special points ($h=\pm J$, $\de = 0$) and $(h=0, \de = \pm 
\ga)$ are shown on the phase boundaries. The spin chain undergoes a quantum
phase transition with $\nu =2$ and $z=1$ when these points are approached 
along the dashed line. On the other hand, the dotted line shows the direction
of the quenching of the transverse field. We quench the system along a gapless
line parallel to the $\ga$-axis (perpendicular to the plane of the paper)
passing through the phase boundary $h^2 = \de^2 + J^2$.} \end{figure}

The minimum energy gap in the excitation spectrum occurs at $k=0$ and $k=
\pi/2$. The corresponding phase boundaries given by $h^2 = \de^2 + J^2$ and
$\de^2 = h^2 + \ga^2$ signal quantum phase transitions from a 
paramagnetic to a ferromagnetic phase and a dimer to ferromagnetic phase
respectively (see Fig. 1). Let us define a new set of Pauli matrices $\tau$ as
$$\tau_i^x = (-1)^i \si_i^x, ~~\tau_i^y = \si_i^y ~~{\rm and}~~
\tau_i^z=(-1)^i \si_i^z,$$ so that the commutation relations of the Pauli 
matrices are preserved. It is interesting to note that under this unitary 
transformation, we arrive at a set of duality relations given by $h \to -
\de$, $\de \to -h$, $J \to -\ga$ and $\ga \to -J$; this signifies 
that the ferro-para transition and the ferro-dimer transition at $~ h^2 = 
\de^2 + J^2$ and $\de^2 = h^2 + \ga^2$~, respectively, are 
essentially identical, both belonging to the quantum Ising universality class
\ct{bunder} with $\nu = z =1$. The phase boundary given by $h^2 = \de^2+J^2$
with $\ga$ arbitrary and $J$ held fixed, defines a critical surface in the 
parameter space spanned by $h$, $\de$ and $\ga$. Similarly, the phase 
boundary $\de^2 = h^2 +\ga^2$ with arbitrary $J$ once again defines 
another critical surface when $\ga$ is held fixed. For $\ga =0$, a 
gapless phase exists with an ordering wave vector $\cos k = \sqrt{h^2 - 
\de^2}/J$ for $\de^2 < h^2 < \de^2 + J^2$. The system undergoes a 
quantum phase transition from a gapless phase to a gapped phase when $|h|$ is 
increased beyond the critical value given by $h_c = \sqrt{\de^2+J^2}$.

The special case with $\de =0$ refers to the well studied anisotropic $XY$ 
spin-1/2 chain in a transverse field \ct{bunder}. In this model, there exists 
an Ising transition line at $h = \pm J$ from the ferromagnetically ordered 
phase to a quantum paramagnetic phase. On the other hand, the vanishing of the
gap at $\ga =0$ signifies another quantum phase transition belonging to a 
different universality class between two ferromagnetically ordered phases.
 
It can be established using a numerical diagonalization of the time-dependent 
Schr\"odinger equation involving the reduced Hamiltonian matrix that when 
the transverse field $h$ or the alternating term $\de$ is quenched as 
$t/\tau$ from $-\infty$ to $\infty$, so that the system crosses the quantum 
critical lines as shown in Fig. 1, the density of defects $n$ in the final 
state satisfies the Kibble-Zurek prediction\ct {viola08}. Our interest however
lies in the generation of defects when the system is quenched along a gapless
line. To achieve such a quenching, we vary the anisotropy parameter $\ga$
linearly as $\ga = t/\tau$ from $-\infty$ to $\infty$, keeping $h$, $\de$
and $J$ fixed in such a way that the system always lies on the phase boundary
$h^2 = \de^2 + J^2$. In the limit $t \to -\infty$, $\ga$ is large and 
negative and hence in the ground state, the expectation value $\bra \si_j^x
\si_{j+1}^x - \si_j^y \si_{j+1}^y \ket = -1$. On the other hand, for 
an adiabatic evolution during the entire period of dynamics, this expectation 
value should be $+1$ in the final state. One can choose different critical 
lines for $\ga$-quenching by choosing different values of $h$ and $\de$ 
on the critical surface.

The eigenvalue $\Lambda_k^-$ given by Eq. (\ref{lam3}) can be written as
\ba &\Lambda_k^-& = [(h-\sqrt{\de^2+J^2 \cos^2 k})^2+ 2h \sqrt{\de^2 + J^2
\cos^2 k} \non \\ 
&+& \ga^2 \sin^2 k -2 \sqrt{h^2\de^2+h^2 J^2 \cos^2 k+\de^2 \ga^2 \sin^2 
k}]^{1/2}. \non \\
& & \ea
On the gapless line $h^2=\de^2+J^2$, the dispersion of the low-energy 
excitations at $k \to 0$ can be approximated as 
\be \Lambda_k^- ~=~ \sqrt{\frac{J^4 k^4}{4(\de^2 + J^2)} + \frac{\ga^2 J^2 
k^2}{\de^2 + J^2}} ~. \label{lam6} \ee

A close inspection of the above excitation spectrum suggests that when 
$\ga$ is quenched along the gapless line, Eq. (\ref{lam6}) can be mapped 
to the spectrum of a $2\times2$ Landau-Zener (LZ) Hamiltonian \ct{landau,sei}
with two linearly approaching time-dependent levels. To show this explicitly,
we note that in the limit of very slow quenching, $\tau \to \infty$, 
defects are produced by sets of modes between which the energy gap 
is very small. For the Hamiltonian in Eq. (\ref{ham1}), this occurs in the 
region of $k=0$ if we take $h^2 = \de^2 + J^2$. Let us first set $\ga =0$ 
and $k=0$. We then see that there are two modes, called $|I \rangle$ and $|II 
\rangle$, whose energies are zero; the other two modes have energies which
are both far from zero and far from each other, and can therefore be ignored
in a slow quenching calculation. The zero energy modes are given by
\ba | I \rangle &=& \frac{1}{\sqrt{\de^2 + (h+J)^2}} ~\left( \begin{array}{c}
\de \\ 0 \\ 0 \\ h+J \end{array} \right), \non \\
| II \rangle &=& \frac{1}{\sqrt{\de^2 + (h+J)^2}} ~\left( \begin{array}{c}
0 \\ \de \\ h+J \\ 0 \end{array} \right). \ea
We now deviate slightly from $k=0$, still keeping $\ga =0$ in Eq. (\ref{ham1}).
Doing degenerate perturbation theory to first order in $Jk^2$, we find that
the modes $|I \rangle$ and $|II \rangle$ remain eigenstates of the Hamiltonian,
but their energies are now given by 
\be E_\pm ~=~ \pm ~\frac{J^2 k^2}{2 \sqrt{\de^2+J^2}}, \ee
where we have used the relation $h^2 = \de^2 + J^2$.
Finally, we introduce the terms involving $\ga$ in Eq. (\ref{ham1}). To
first order in $\ga$, we find that the Hamiltonian in the basis 
of $|I \rangle$ and $|II \rangle$ is given by
\be h_k ~=~ \frac{1}{\sqrt{\de^2 + J^2}} ~\left( \begin{array}{cc}
J^2 k^2 /2 & -i \ga J k \\ i \ga J k & -J^2 k^2 /2 \end{array} \right). \ee
If we now perform a unitary transformation and vary $\ga$ in time, we see 
that the Hamiltonian is of the LZ form,
\be h_k ~=~ \left[ \begin{array}{cc} {\tilde \ga(t)} k & {\tilde J}^2 k^2 /2 \\
{\tilde J}^2 k^2 /2 & {-\tilde \ga(t)}k \end{array} \right], \label{ham2} \ee
where $\tilde \ga$ and $\tilde J$ are renormalized parameters given by $\tilde
\ga = \ga J/\sqrt{\de^2+J^2}$ and $\tilde J^2 = J^2/ \sqrt{\de^2+J^2}$. The 
diagonal terms in Eq. (\ref{ham2}) describe two time-dependent levels 
approaching each other linearly in time (since $\ga = t/\tau$), while the 
minimum gap is given by the off-diagonal term ${\tilde J}^2 k^2 /2$. The 
probability of excitations $p_k$ from the ground state to the excited state 
for the $k$-th mode is given by the Landau-Zener transition formula 
\ct{landau,sei} 
\be p_k ~=~ \exp ~[-~\frac{2\pi \tilde J^4k^4}{8 k \frac{d\tilde \ga(t)}{dt}}]
=\exp~[-~\frac{\pi J^3k^3\tau}{4\sqrt{\delta^2+J^2}}]. \label{pk} \ee

\noindent Note that for large $\tau$, $p_k$ is dominated by values of $k$
close to 0. The density of excitations $n$ in the final state and in the large
$\tau$ limit is obtained by integrating over all modes in Eq. (\ref{pk}), 
\be n ~=~ \frac{2}{\pi} ~\int_0^{\pi/2} dk ~p_k ~ \sim ~\frac{1}{\tau^{1/3}}.
\label{n} \ee
The numerical integration of Eq. (\ref{n}) for $\de = J = 1$ is shown in 
Fig. 2 (a); although we have used the expression given in Eq. (\ref{pk})
for all values of $k$, the dominant contribution to $n$ comes from the
region near $k=0$ where (\ref{pk}) can be trusted.
This shows that when quenched along a gapless line, the density 
of defects in the final state exhibits a slower decay with $\tau$ as compared
to the $1/{\sqrt \tau}$ which is observed in the case when the gapless line is
crossed by varying $h$ \ct{viola08}. Our study also reveals that although the
spectrum of this system involves four levels, the quenching dynamics along the
gapless line is essentially a two-level problem which can be studied using
an effective Landau-Zener theory with parameters renormalized by $\de$. 

\begin{figure}
\includegraphics[height=3.5in]{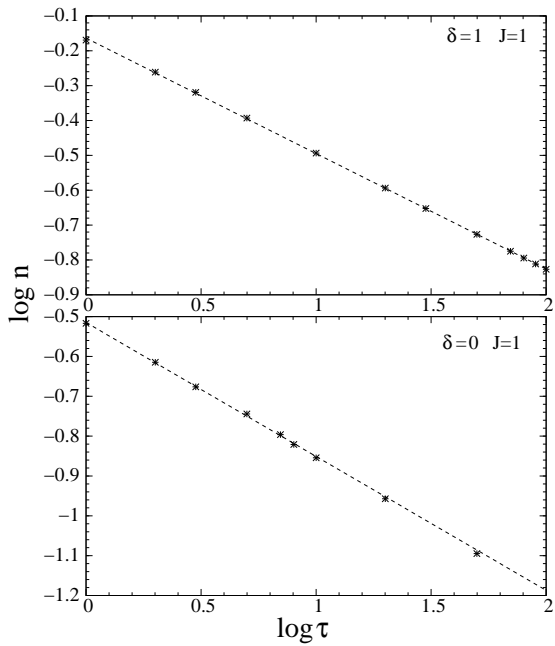}
\caption{Fig. 2 (a) shows the variation of defect density $n$ with $\tau$ for
$\de = J = 1$ obtained by numerically integrating Eq. (\ref{n}), using $p_k$
given in Eq. (12). Fig. 2 (b) shows $n$ obtained by direct numerical 
integration of the Schr\"odinger Equation with $\de =0$. In both figures,
the fitted data shows a slope of $-1/3$ confirming the $n \sim 1/\tau^{1/3}$
behavior.} \end{figure}

The case $\de=0$ corresponds to quenching the system along the Ising 
critical line of the transverse $XY$ model, and one obtains an identical 
scaling of the defect density. Fig. 2 (b) shows the $1/3$ power-law obtained
by numerically solving the Schr\"odinger equation for the anisotropic
transverse $XY$ model when the anisotropy parameter $\gamma$ is quenched
along the gapless line $h=-J=1$. (The Hamiltonian in Eq. (\ref{ham1}) 
has a $2 \times 2$ block diagonal form if $\de =0$).
One can also propose an alternative quenching
scheme where the strength $J$ is quenched from $-\infty$ to $\infty$,
keeping $h$, $\de$ and $\ga$ constant with $\de^2 = h^2 + \ga^2$.
The duality relation discussed above leads to the conclusion that this 
quenching scheme is equivalent to the previous one with $\ga = t/\tau$,
and it yields a similar $1/\tau^{1/3}$ behavior. 

The $XY$ chain with an alternating transverse field can be further generalized
by incorporating additional alternations in the strength or in the anisotropy 
of the interaction with the period of alternation being two. We denote the 
alternation in the strength and the anisotropy by $J_s$ and $\ga_s$, 
respectively, and for simplicity choose $J_s=0$. Using a similar Jordan-Wigner
transformation (Eq. (\ref{jw})) followed by a Fourier transformation, we find 
an excitation spectrum of the form
\ba &\Lambda_k^\pm & = [ h^2 + \de^2 + (J^2+\ga_s^2) \cos^2 k + \ga^2 \sin^2 k
\non \\
&-& 2 \sqrt{(h^2 + \ga_s^2 \cos^2 k)(\de^2 + J^2 \cos^2 k) + \de^2 \ga^2
\sin^2 k} ]^{1/2}, \non \\
& & \label{lam8} \ea
where the eigenvalue $\Lambda_k^-$ has to be analyzed to explore the quenching
dynamics. Eq. (\ref{lam8}) shows that the role of the alternation $\ga_s$ is
to renormalize the strength $J$; consequently, the phase boundary separating 
the paramagnetic and the ferromagnetic phase gets shifted to $h^2 = \de^2 +
J^2 - \ga_s^2$, with arbitrary $\ga$, at which the gapless excitations occur 
at $k=0$. Quenching $\ga$ as $t/\tau$ along the new phase boundary $h^2 = 
\de^2 + J^2 - \ga_s^2$ with fixed values of $J$, $h$ and $\ga_s$, once again 
takes the system along a gapless line on a critical surface.

The dynamics can be reduced to a two-level problem as before, and the 
low-energy excitations above the gapless line are given by
\be \Lambda_k^- ~=~ \sqrt{\frac{(J^2 - \ga_s^2)^2 k^4}{4(\de^2 + J^2)}~
+~ \frac{\ga^2 J^2 k^2}{\de^2 + J^2}}~. \ee
The defect density $n$ in the final state decreases with $\tau$ as $1/
\tau^{1/3}$ as can be derived from the Landau-Zener formula in an identical
fashion. Similarly, one may set $\ga_s =0$ and $J_s \neq 0$ and consider an 
equivalent quenching scheme resulting in an identical scaling of the defect 
density. 

The behavior of the defect density when quenched
along a gapless line suggests the following general 
scaling relation of the defect density for a $d$-dimensional quantum system. 
Let the excitations on the gapless quantum critical line be of the form 
$\omega_{\vec k} \sim \alpha | \vec k|^z$, where $z$ is the dynamical exponent
and the parameter $\alpha=t/\tau$ is quenched from $-\infty$ to $\infty$. 
Using a perturbative method involving the Fermi Golden rule along with the 
fact that the system is initially prepared in the ground state, the defect 
density can be approximated as \cite{polkovnikov07}
\be n ~\simeq~ \int \frac {d^d k}{(2\pi)^d} ~|\int_{-\infty}^\infty d\alpha ~
\langle \vec k | \frac {\partial}{\partial \alpha} |0 \rangle ~e^{i\tau 
\int^{\alpha} \de \omega_{\vec k} (\alpha') d\alpha'}|^2 . \ee
Assuming a general scaling form of the instantaneous excitation $\de 
\omega_{\vec k} (\alpha') = k^a f(\frac {\alpha k^z}{k^a})$, where $k = |
\vec k|$, and $k^a$ denotes the higher order term in the excitation spectrum 
on the gapless line. Defining a new variable $\xi = \alpha k^{z-a}$, we 
obtain the scaling behavior of the defect density as 
\be n ~\sim ~1/\tau^{d/(2a -z)} ~. \label{general} \ee
The case $d=1$, $a =2$ and $z=1$ has been discussed in the present work.
Note that the correlation length exponent $\nu$ does not appear in the 
expression in Eq. (\ref{general}) because our quench dynamics always
keeps the system on a critical line.

\section{conclusion}

In conclusion, we have studied the defect density produced in the final state 
when a generalized spin-1/2 $XY$ chain with an alternating transverse field 
as well as an alternating nearest-neighbor interaction is quenched along the 
Ising critical line by varying the anisotropy parameter $\gamma$. We show that
the non-adiabatic transition probability and hence the defect density can be 
estimated using an equivalent Landau-Zener problem in which the parameters are
renormalized by the alternating parameters $\delta$ and $\gamma_s$ (or $J_s$).
We find that the defect density decays with the characteristic time scale
of quenching given by $1/\tau^{1/3}$. The defect scaling exponent obtained
here does not fit the KZ scaling $1/\tau^{d \nu/(z \nu + 1)}$. In the present
quenching scheme, the system is always on a critical surface, and therefore
the critical exponent $\nu$ does not appear in the scaling of the defect
density. The quenching scheme used here is different from the other quenching
schemes through a gapless phase \ct{polkovnikov07,sengupta08,santoro08} where
the system starts from a non-critical (gapped) point, goes through a critical
(gapless) point or critical surface, and eventually ends again at a 
non-critical point.

\begin{center}
{\bf Acknowledgments}
\end{center}

U.D. and A.D. acknowledge Victor Mukherjee for collaboration in related works,
and Satyajit Banerjee and Debashish Chowdhury for interesting discussions.
D.S. thanks Krishnendu Sengupta for discussions and DST, India for financial
support under Project No. SR/S2/CMP-27/2006.

\end{document}